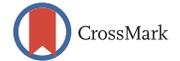

## Research Article
## Egg Shell Quality and Bone Status as Affected by Environmental Temperature, Ca and Non-Phytate P Intake and *in vitro* Limestone Solubility in Single-Comb White Leghorn Hens

[1]Bingfan Zhang, [2]Jordan Weil, [2]Antonio Beita Guerra, [2]Pramir Maharjan, [2]Katie Hilton, [2]Nawin Suesuttajit, [2]Diego Martinez Patino Patr and [2]Craig N. Coon

[1]802 Jinchengyuan, Fuhao Garden, 50 Fusong Road, Caoyang District, Changchun, Jilin, PRC 130012, China
[2]Center of Excellence for Poultry Science, University of Arkansas, Fayetteville, North Carolina, AR 72701, United States

## Abstract

**Background and Objective:** Environmental temperature (ET) often changes the nutrient intake/output for layers. Changing feed formulations based on ET may need to be utilized to obtain optimum performance, shell quality and bone status. This study was conducted to investigate the effects of temperature, Ca intake, non-phytate P (NPP) intake and *in vitro* limestone solubility (LS) on egg-shell quality and bone status in commercial White Leghorn hens. **Materials and Methods:** Two 9 weeks feeding experiments were conducted with two ages of commercial White Leghorns housed in either a constant (21.1°C) thermoneutral temperature or in hot ambient temperatures (constant: 26.6°C or 32.2°C or cycling: 01:00-09:00 h at 26.7°C; 09:00-13:00 h at 29.4°C; 13:00-18:00 h at 35.0°C and 18:00-01:00 h at 29.4°C with 24 h mean of 29.7°C). In Experiment (EXP) 1, 360-130 week molted hens were housed in a constant ET at 21.1, 26.6 and 32.2°C and fed 3 levels of Ca (3.8 g, 4.9 g and 6 g h$^{-1}$ day$^{-1}$), 2 levels of NPP (450 and 600 mg h$^{-1}$ day) while feeding 2 different LS (*in vitro* solubility of 34.1 and 48.4%). In EXP 2, 480-48 week old hens were housed in constant ET at 21.1°C or in cycling hot ET(with 24 h mean of 29.7°C: 01:00-09:00 h at 26.7°C; 09:00-13:00 h at 29.4°C; 13:00-18:00 h at 35.0°C and 18:00-01:00 h at 29.4°C) and fed 2 different LS (*in vitro* solubility of 34.1 and 48.4%) and four predicted intakes of NPP ( 250, 350, 450 and 550 mg h$^{-1}$ day$^{-1}$). **Results:** Egg mass and shell weight per unit surface area (SWUSA) decreased with increasing ET (p<0.05), especially when ET was 29.7°C (cycling mean ET)or a constant ET was 32.2°C. Feeding layers a low soluble larger particle size limestone instead of a highly soluble limestone produced beneficial effects for SWUSA at the thermoneutral ET (21.1°C) but the beneficial effect was less or disappeared when ET was ≥26.6°C in EXP 1 and 2. Feeding layers 245 and 353 mg NPP h$^{-1}$ day$^{-1}$ supported satisfactory bone status at 21.1°C, however layers housed at ≥30°C needed an additional intake of 50 mg NPP h$^{-1}$ day$^{-1}$ to support bone integrity. Results of EXP 1 and 2 indicates that ≥48 week old layers housed in thermoneutral or warmer ET require a minimum of ≥4.2 g Ca h$^{-1}$ day$^{-1}$ for maintaining optimum shell quality and bone integrity. **Conclusion:** Feeding low LS (34.1% *in vitro* solubility) improved egg shell quality only for hens housed in thermoneutral ET (21.1°C) and did not improve egg shell quality at higher ET (constant or cycling). Daily NPP intake of 245 and 353 mg h$^{-1}$ day$^{-1}$ supported optimum egg production and bone status at 21.1°C, respectively. A higher NPP and Ca intake may be required for bone status compared to egg production, especially in older hens.

## INTRODUCTION

Large variations exist in the literature on the recommended Ca and P dietary requirement for laying hens[1]. The Ca requirement was reduced from a previous suggested intake of 3.75[2] to 3.25 g h$^{-1}$ day$^{-1}$ by NRC[3] after decades of consistently increasing the Ca intake requirement. The available P or non-phytate P (NPP) requirement for commercial layers was also reduced to 250 mg h$^{-1}$ day$^{-1}$. A review by Scheideler[4] suggested that White-Egg primary breeders require a daily Ca intake of 3.5-4.1 g during peak of production, 3.75-4.3 g at post peak and 3.9-4.4 g during the late stages of the laying cycle and non-phytate P (NPP) intake of 420-500 mg h$^{-1}$ day$^{-1}$ at peak of lay, 380-480 mg h$^{-1}$ day$^{-1}$ at post peak and 300-450 mg h$^{-1}$ day$^{-1}$ during the late stages of the laying cycle. Rostagno *et al*.[5] suggested the Ca and NPP requirements for White-Egg layers producing 52 g egg mass day$^{-1}$ housed in thermoneutral or 21-31°C environmental temperatures is 4.2 g and 320mg per day, respectively. The committee for FEDNA[6] suggested the Ca and NPP (digestible P) requirement for commercial light layers in cages consuming between 111-117g feed day$^{-1}$ from 26-50 and > 50 weeks is 3.85-4.0% Ca, 0.31-0.33% digestible P and 3.9-4.2% Ca, 0.29-0.32% digestible P.

Reducing dietary P as hen's age has been shown, in general, to have no adverse effects on egg production[1,7-9] although a severe P deficiency has been shown to depress egg production and increase mortality[10-12]. Phase-feeding dietary P for layers during the production period with decreasing NPP levels such as 0.46, 0.36 and 0.26% or even a lower decreasing NPP series with 0.36, 0.26 and 0.16% available P from 24-39, 39-51 and 51-71 week, respectively, has been reported to support high rates of egg production without excessive depletion of body P[13]. Phase-feeding lower levels of available P such as 0.34, 0.25 and 0.15% for 24-39, 39-51 and 51-71 week periods had no adverse affects on performance at thermoneutral temperature, however, the researchers observed the lower levels of available P diets (0.25 and 0.15%) provided an available P intake of 250 mg and 150 mg, respectively and caused a significant reduction of layer body P. The layers were fed the low available P diets in hot environmental temperatures (ET) which may have exacerbated the body P reduction[14]. Several researchers have reported possible increased P requirements when hens were housed in a hot ET[8,15,16]. Belay *et al*.[17] has reported that broilers housed in a constant hot ET show ≈2x increase in urine output because of increased water intake and ≈2x loss of calcium and P. In addition to environmental temperature having an effect on retention of Ca and P, the apparent digestibility, utilization and retention of dietary P and Ca for light commercial layers[18,19] and heavy broiler breeders[20,21] has been reported to be dependent upon the particle size of Ca in the diet. The research reported herein was conducted to determine the effects of dietary calcium intake, P intake and limestone solubility (LS)/particle size on layer bone status and egg shell quality for layers housed in normal thermoneutral temperatures and adverse hot ET.

## MATERIALS AND METHODS

**Experiment 1 (EXP 1):** A total of 360 Single-Combed White-Egg Leghorn hens (H&N), 130 week of age, molted twice, were randomly allocated one bird per cage (25.4×35.6 cm) with individual feeders and drinkers in three environmentally controlled rooms (120 hens per room). The environmental temperature (ET) for each of the three rooms was initially adjusted at 21.1°C with relative humidity (RH) maintained between 40-50%. The ET was increased by 2.0°C per week for two of the rooms until a constant temperature of 26.6 and 32.2°C was achieved providing 21.1, 26.6 and 32.2°C for the three rooms. The layers were acclimated for one week at the three constant temperatures. Following acclimation, the layers were fed a pre-study diet (Table 1) for a two-9 week period for the purpose of determining feed consumption of the hens housed in different ET. The diet used in the pre-study was a corn-soy diet which was formulated to contain 2950 kcal kg$^{-1}$, 17% protein, 4% Ca and 0.42% NPP. Test hens with a minimum of ≥75% egg production (EP) were used in the calculation of the average feed consumption in each of the temperatures. The concentrations of dietary Ca, NPP and other nutrients were adjusted according to temperature in the formulation of the experimental diets based on the ME intakes determined in the pre-study feeding period. The final experimental diets (Table 2) were formulated to ensure that the hens housed in different temperatures would consume an equal amount of nutrients other than energy when the hens have the same egg production. This is based on the theory that the feed intake for the mature layer is regulated primarily by energy intake. The average daily feed intake for the hens determined in the pre-study feeding period was 119.8, 117.3 and 111.3 g h$^{-1}$ day$^{-1}$ for the 21.1, 26.6 and 32.2°C ET treatments, respectively. The Ca and NPP provided in experimental diets was supplied only from corn, soybean meal, limestone and dicalcium P to eliminate the issue of Ca and NPP quantity provided by a commercial exogenous phytase.

Table 1: Composition of the pre-environmental test diet for experiment 1 (EXP 1) and experiment 2 (EXP 2)

| Ingredients | Diet (%) |
|---|---|
| Corn | 58.28 |
| Soybean meal, 47% crude protein | 25.47 |
| DL-methionine(98.5%) | 0.14 |
| Limestone Unical S[1] | 4.62 |
| Limestone Shell and Bone Blend[1] | 4.62 |
| Salt | 0.25 |
| Vitamin mixture[2] | 0.04 |
| Mineral mixture[3] | 0.04 |
| Choline chloride, 60% | 0.13 |
| Poultry fat | 4.37 |
| Dicalcium phosphate | 1.73 |
| Sodium bicarbonate | 0.31 |
| **Calculated composition** | |
| ME (kcal kg$^{-1}$) | 2950.00 |
| Crude protein (%) | 17.00 |
| Non-phytate P (%) | 0.42 |
| Calcium (%) | 4.00 |

[1]ILC Shell and bone blend (SBB) (2856 microns, *in vitro* solubility of 34.1%) and ILC Unical S (450 microns, *in vitro* solubility of 48.4%). [2]Vitamin mixture provides in milligrams per kilogram of diet; Vitamin A: 5,500 IU, Vitamin E: 25 IU, Menadione: 1.45 mg, Cholecalciferol: 1,100 IU, Riboflavin: 5.4 mg, Pantothenic acid: 23 mg, Nicotinic acid: 55 mg, Vitamin B12: 9.9 mg, Vitamin B6: 9.5 mg, Thiamine: 5.4 mg, Folic acid: 1.8 mg, Biotin: 0.28 mg. [3]Trace mineral mixture provides in milligrams per kilogram of diet; Mn: 68, Zn: 61, Fe: 120, Cu: 7, I: 0.7, Se: 0.3

Table 2: Diets for experiment 1(EXP 1)[1]

| Ingredients | 21.1°C | | | 26.6°C | | | 32.2°C | | |
|---|---|---|---|---|---|---|---|---|---|
| | 3.8 g Ca | 4.9 g Ca | 6 g Ca | 3.8 g Ca | 4.9 g Ca | 6 g Ca | 3.8 g Ca | 4.9 g Ca | 6 g Ca |
| Corn (ground) | 63.31 | 58.49 | 53.78 | 61.85 | 56.92 | 51.98 | 58.05 | 52.85 | 47.72 |
| Soybean meal (47%CP) | 23.17 | 24.02 | 24.85 | 24.16 | 25.03 | 25.9 | 26.72 | 27.64 | 28.54 |
| Poultry fat | 3.07 | 4.7 | 6.3 | 3.35 | 5.03 | 6.7 | 4.11 | 5.87 | 7.62 |
| Limestone (SBB)[2] | 6.61-7.03 | 8.94-9.36 | 11.29-11.71 | 6.74-7.17 | 9.13-9.56 | 11.51-11.94 | 7.54 | 9.6-10.05 | 12.12-12.57 |
| Limestone (Unical S)[3] | 6.67-7.09 | 9.03-9.45 | 11.39-11.82 | 6.81-7.24 | 9.21-9.65 | 11.62-12.06 | 7.61 | 9.69-10.15 | 12.23-12.69 |
| Dicalcium phosphate[4] | 1.49-2.17 | 1.50-2.19 | 1.52-2.2 | 1.53-2.23 | 1.54-2.24 | 1.56-2.26 | 1.63 | 1.65-2.38 | 1.66-2.4 |
| Silica[5] | 0.7-1.0 | 0.68-1.0 | 0.59-1.0 | 0.70-1.0 | 0.67-1.0 | 0.64-1.0 | 0.7-1.0 | 0.66-1.0 | 0.58-1.0 |
| Sodium bicarbonate | 0.25 | 0.25 | 0.25 | 0.32 | 0.31 | 0.30 | 0.32 | 0.31 | 0.25 |
| Choline Cl (60%) | 0.13 | 0.13 | 0.13 | 0.13 | 0.13 | 0.13 | 0.13 | 0.13 | 0.13 |
| DL-methionine (98.5%) | 0.12 | 0.13 | 0.13 | 0.13 | 0.13 | 0.14 | 0.14 | 0.15 | 0.16 |
| Lysine, HCl, H$_2$O | 0.03 | 0.01 | | 0.03 | 0.01 | | 0.03 | 0.01 | |
| Salt | 0.3 | 0.28 | 0.25 | 0.24 | 0.25 | 0.26 | 0.24 | 0.25 | 0.28 |
| Vitamin premix[6] | 0.04 | 0.04 | 0.04 | 0.04 | 0.04 | 0.04 | 0.04 | 0.04 | 0.04 |
| Trace mineral premix[7] | 0.04 | 0.04 | 0.04 | 0.04 | 0.04 | 0.04 | 0.04 | 0.04 | 0.04 |
| **Calculated analysis** | | | | | | | | | |
| AME (kcal kg$^{-1}$) | 2950 | 2950 | 2950 | 2950 | 2950 | 2950 | 2950 | 2950 | 2950 |
| Crude protein (%) | 16.33 | 16.33 | 16.33 | 16.68 | 16.68 | 16.68 | 17.58 | 17.58 | 17.58 |
| Dig. TSAA (%) | 0.59 | 0.59 | 0.59 | 0.61 | 0.61 | 0.61 | 0.64 | 0.64 | 0.64 |
| Dig lysine (%) | 0.74 | 0.74 | 0.74 | 0.76 | 0.76 | 0.76 | 0.80 | 0.80 | 0.80 |
| Non-phytate P (%) | 0.38-0.50 | 0.38-0.50 | 0.38-0.50 | 0.38-0.51 | 0.38-0.51 | 0.38-0.51 | 0.40-0.54 | 0.40-0.54 | 0.40-0.54 |
| Calcium (%) | 3.17 | 4.09 | 5.01 | 3.17 | 4.18 | 5.12 | 3.41 | 4.40 | 5.39 |

[1]Constant environmental temperatures of 21.1, 26.6 and 32.2°C. The test diets were formulated for feed intakes of 119.8, 117.3 and 111.3 g, respectively, for the three different environmental temperatures. A basal was made for each calcium intake for each temperature and only one of the limestones was added for each test diet and only one level of dicalcium phosphate was added. [2]ILC Shell and Bone Blend, (SBB), 2856 microns, 34.1% *in vitro* soluble, analyzed Ca, 39.08%. [3]ILC Unical S, 450 microns, 48.4% *in vitro* soluble, analyzed Ca, 38.71%. [4]Dicalcium phosphate was added to each basal to provide a projected daily intake of 450 and 600 mg NPP. [5]White washed sand; added to adjust the experimental diet to 100% for the two particle sizes of limestone and the added dicalcium phosphate. [6]Vitamin mixture provides in milligrams per kilogram of diet; Vitamin A: 5,500 IU, Vitamin E: 25 IU, Menadione: 1.45 mg, Cholecalciferol: 1,100 IU, Riboflavin: 5.4 mg, Pantothenic acid: 23 mg, Nicotinic acid: 55 mg, Vitamin B12: 9.9 mg, Vitamin B6: 9.5 mg, Thiamine: 5.4 mg, Folic acid: 1.8 mg, Biotin: 0.28 mg. [7]Trace mineral mixture provides in milligrams per kilogram of diet; Mn: 68, Zn: 61, Fe: 120, Cu: 7, I: 0.7, Se: 0.3

Hens were randomly assigned into a 3×2×3×2 factorial arrangement of treatments with 10 replicates of individually caged layers. Three constant ET, two different LS (*in vitro* solubility: 34.1 and 48.4%), three predicted dietary Ca intakes (3.8, 4.9 and 6.0 g day$^{-1}$ h$^{-1}$) and two predicted NPP intake levels (450 and 600 mg day$^{-1}$ h$^{-1}$) were used in the experiment.

**Experiment 2(EXP2):** A total of 480 Single-Combed White-Egg Leghorn hens (Dekalb-XL), 48 week of age, were randomly allocated to four environmentally controlled rooms (120 hens in each of the rooms). The temperature of the environmental rooms was initially adjusted at 21.1°C. The RH of the rooms were maintained between 40-50% for all temperatures. Two rooms were assigned for each of the two temperature regimes: thermoneutral (21.1°C) constant ET or warmer cycling ET regime (01:00-09:00 hr at 26.7°C; 09:00-13:00 h at 29.4°C; 13:00-18:00 h at 35.0°C and 18:00-01:00 h at 29.4°C with a 24 h mean of 29.7°C). The cycling ET regime was not initiated until 29.4°C was reached with a weekly 2.0°C increment. After two weeks acclimation in the thermoneutral

Table 3: Diets for experiment 2 (EXP 2)[1]

| Ingredients | 21.1°C constant (%) | Cycling hot ET (%) |
|---|---|---|
| Corn, ground (8.4% CP) | 56.47 | 34.98 |
| Soybean meal (47% CP) | 27.78 | 42.22 |
| DL-methionine (98.5%) | 0.15 | 0.24 |
| Limestone SBB[2] | 7.85-8.78 | 9.97-11.15 |
| Limestone Unical S[3] | 7.93-8.86 | 10.07-11.26 |
| Dicalcium phosphate[4] | 0.67-2.18 | 0.97-2.90 |
| Silica[5] | 0.32-1.0 | 0.14-1.0 |
| Sodium Bicarbonate | 0.25 | 0.25 |
| Salt | 0.27 | 0.28 |
| Vitamin mixture[6] | 0.04 | 0.04 |
| Mineral mixture[7] | 0.04 | 0.04 |
| Choline chloride (60%) | 0.13 | 0.13 |
| Poultry fat | 4.42 | 8.70 |
| **Calculated composition** | | |
| ME (kcal kg$^{-1}$) | 2950 | 2950 |
| Crude protein (%) | 17.95 | 23.02 |
| Digestible TSAA (%) | 0.65 | 0.84 |
| Digestible Lysine (%) | 0.82 | 1.16 |
| Non-Phytate P (%) | 0.23-0.51 | 0.29-0.65 |
| Calcium (%) | 3.67 | 4.71 |

[1]Constant 21.1°C and Cycling hot ET: Cycling environmental temperatures with 24 h means of 29.7°C (01:00-09:00 h at 26.7°C; 09:00-13:00 h at 29.4°C; 13:00-18:00 h at 35.0°C and 18:00-01:00 h at 29.4°C). [2]ILC Shell and Bone Blend, (SBB), 2856 microns, 34.1% *in vitro* soluble, analyzed Ca, 39.08%. [3]ILC Unical S, 450 microns, 48.4% *in vitro* soluble, analyzed Ca, 38.71%. [4]Dicalcium phosphate was added to each basal to provide a projected intake of 250, 350, 450 and 550 mg NPP [5]White washed sand; added to adjust the experimental diet to 100% for the two particle sizes of limestone and the added dicalcium phosphate [6]Vitamin mixture provides in milligrams per kilogram of diet; Vitamin A: 5,500 IU, Vitamin E: 25 IU, Menadione: 1.45 mg, Cholecalciferol: 1,100 IU, Riboflavin: 5.4 mg, Pantothenic acid: 23 mg, Nicotinic acid: 55 mg, Vitamin B12: 9.9 mg, Vitamin B6: 9.5 mg, Thiamine: 5.4 mg, Folic acid: 1.8 mg, Biotin: 0.28 mg. [7]Trace mineral mixture provides in milligrams per kilogram of diet; Mn: 68, Zn: 61, Fe: 120, Cu: 7, I: 0.7, Se: 0.3

and warmer cycling ET, a two-week pre-study was conducted to determine ME intakes for the two different temperature regimes. The layers were fed the same pre-study diet (Table 1) used in EXP 1 and the feed intake during the pre-study period was then used to determine the concentrations of dietary Ca, NPP and other nutrients that would be needed in the formulation of the experimental diets. The final experimental diets (Table 3) were formulated to ensure that the hens with the same egg mass output would consume equal amount of nutrients other than energy when housed in different ET.

Hens were randomly assigned into a 2×2×4 factorial arrangement of treatments with 30 replicates of each of the treatments. The treatments for EXP 2 consisted of a two temperature regimen, two LS (34.1 and 48.4%) and four predicted dietary NPP intake levels (250, 350, 450 and 550 mg h$^{-1}$ day$^{-1}$). The predicted Ca intake was fixed at 4 g h$^{-1}$ day$^{-1}$ for all the treatments.

**Limestone solubility (LS):** The solubility value of each particle size of limestone was determined by a modified Minnesota Percentage Weight Loss Method[22,23]. The modified method[23] differed from the original procedure in the acidity and volume of the solution used (200 mL of 0.2 N HCl in the modified vs 100 mL of 0.1 N HCl in the original method). Limestone used in EXP 1 and 2 was from the same commercial source (ILC Resources, Des Moines, Iowa) and was screened into two different sizes. The mean size of the Shell and Bone Builder Blend™ larger particle was 2856 microns and the smaller particle Unical-S™ limestone product had a mean particle size of 450 microns. The limestone particle size was measured by laser diffraction instead of sieve screen analysis. Solubility was the basis of evaluating limestone quality in present research because of previous research showing solubility produced a higher correlation between shell quality and bone status compared to particle size measurements[24].

**Poultry husbandry, production variables, shell quality and feed Ca and NPP:** Feed and water were provided *ad libitum* and birds were exposed to 16 h of light day$^{-1}$ throughout the 9 week experiment (one week of acclimation, two weeks of pre-study and 6 weeks of experimental treatment). Daily egg production and biweekly feed consumption were recorded. Eggs laid on 3 consecutive days by each hen were collected and weighed every week. The shell weight and shell weight per unit of surface area (SWUSA)[25] of the collected eggs were determined. The Ca and NPP contents of the basal diets and limestone was analyzed by an atomic absorption spectrophotometer and the photometric method[26] before experimental diets were mixed. Each hen was weighed at the beginning and the end of the six-wk experimental treatment.

**Bone breaking force (BBF) and bone ash concentration (BAC):** The hens were euthanized (all hens in EXP 1 and 8 hens from each of the treatments in EXP 2) by cervical dislocation, right tibia bones were removed and stored at -20°C until tested for the various bone parameters. The bone volume was taken by weight change in water method[27]. Briefly, tibia bones were weighed in the air and in the water. The weight change equals the weight of water replaced by the bone. Bone ash weight was obtained after ashing at 600°C for 24 h. BAC was calculated by dividing bone ash weight by the bone volume. BBF was measured by an Instron Testing Machine (Model 1122; Canton, MA 02021). Tibia bones were supported by a fulcrum with 8.5 cm width. A probe with 1.4 cm length and 0.3 cm at the base was attached to a 500 kg load cell with a crosshead speed of 200 mm min$^{-1}$.

**Statistics:** Data was analyzed by general linear models (GLM) and regression procedures using statistical analysis software

(SAS)[28]. A completely randomized design was used. Data was analyzed using ANOVA and differences among the means were tested using Duncan's multiple range tests at 5% level of significance. Superscripts were used to show statistical differences. The average measurements for each variable were used to analyze data. Hens that were going through natural molting during the experiments were excluded from the analysis. All procedures regarding the use of live animals in this study were carried out in accordance with the Animal Use Protocol 03008, which was approved by the University of Arkansas Institutional Animal Care and Use Committee.

## RESULTS AND DISCUSSION

Hen day percent egg production (HDEP), egg weight (EW) and egg mass (EM) were significantly (p<0.01) decreased by temperature increment in both experiments. However, there were no significant reductions in performance caused by increases in ET for hens housed at 26.6°C compared to hens housed at 21.1°C in EXP 1 (Table 4). The decrease in EW in present study with increasing ET has been reported earlier[29]. The results also support the earlier findings that egg output falls more rapidly when temperature is above 29°C and layers start to pant[30,31].

Hens housed at 32.2°C consumed less feed than layers housed at 21.1 and 26.6°C rooms (EXP 1). Although, hens housed at 26.6°C produced daily about 1.2 g more EM output, the hens consumed approximately 4 g h$^{-1}$ day$^{-1}$ less feed when compared with hens housed at 21.1°C environmental room (Table 4). The hens housed at 26.6°C showed a more efficient conversion of the diet into EM compared to hens housed at 21.1°C. Although, the efficient conversion was reflected by the higher ratio of EM to feed for hens housed at 26.6°C compared to hens housed at 21.1°C, the difference between the two temperature treatments was not significant (Table 4). The reduction in feed intake (FI) with environmental temperature has been well documented in earlier studies[29,32-37]. The same trend was found in EXP2, in which lower FI and better feed conversion were accompanied with higher ET (Table 4). The ratio of EM to feed was significantly impaired for hens housed in a constant temperature at 32.2°C in EXP 1. Metabolizable energy efficiency for EM output has been reported to improve for layers housed at 33.9°C compared to 23.9°C due to the lower energy maintenance requirement with higher ET when equal EM was achieved[36-38]. The lower feed efficiency at 32.2°C in EXP 1 is due to the large drop in EM production caused by a severe reduction of 12.2 g day$^{-1}$ in FI compared to hens housed at 26.6°C and reduction of 16.2 g day$^{-1}$ in FI compared to hens housed at 21.1°C. The HDEP for hens in EXP 1 housed at 32.2°C was significantly lower than that of layers housed at 21.1°C.

Egg shell quality (SWUSA) decreased with increasing ET in both EXP 1 and EXP 2. However no significant difference was found in SWUSA between 26.6 and 32.2°C groups (Table 4). The negative effect of high ET on egg shell quality has been documented in numerous reports[29,33,35,39,40].

In contrast to shell quality, both BBF and BAC at 32.2°C were significantly higher than BBF and BAC in layers housed in the 2 lower constant ET in EXP 1. BBF and BAC did not increase for layers housed in the hot cyclic ET in EXP 2 (Table 4). This may indicate that the increase in BBF and BAC at 32.2°C in EXP 1 was due to the severely depressed HDEP at the constant hot ET thus eliminating the quantity of Ca and P being mobilized from medullary bone during egg shell formation.

The actual intakes (AI) of Ca and NPP in different treatments are summarized in Table 5 for EXP 1 and Table 6 for EXP 2. The 3.8, 4.9 and 6.0 g predicted Ca intakes in EXP 1 were 3.4, 4.4 and 5.4 g for AI (Table 5). The predicted intakes instead of the AI's are used in the discussion for the sake of simplicity. The SWUSA was lower for the hens consuming 3.8 g Ca h$^{-1}$ day$^{-1}$ compared to hens consuming 4.9 and 6.0 g Ca h$^{-1}$ day$^{-1}$ in EXP 1 across ET (Table 7). There was no significant difference in SWUSA between the two higher dietary Ca groups (4.9 and 6.0 g h$^{-1}$ day$^{-1}$). Similar results were found for the bone parameters. BBF and BAC were significantly lower for the 3.8 g Ca group compared to the

Table 4: Effect of environmental temperature on performance, shell quality and bone parameters

| | Temperature (°C) | Egg production (%) | Egg weight (g) | Egg mass (g h$^{-1}$ day$^{-1}$) | Feed intake (g h$^{-1}$ day$^{-1}$) | Egg mass: feed | BBF[1] (kg) | BAC[2] (mg/cm$^3$) | SWUSA[3] (mg/cm$^3$) |
|---|---|---|---|---|---|---|---|---|---|
| EXP 1 | 21.1 | 74.200[a] | 67.000[a] | 48.500[a] | 113.800[a] | 0.427[a] | 10.910[b] | 416.900[b] | 75.100[a] |
| | 26.6 | 76.700[a] | 65.600[a] | 49.700[a] | 109.800[a] | 0.453[a] | 11.170[b] | 435.200[b] | 73.000[b] |
| | 32.2 | 59.100[b] | 63.200[b] | 34.300[b] | 97.600[b] | 0.351[b] | 11.850[a] | 462.400[a] | 72.900[b] |
| | Pooled SE | 1.254 | 0.275 | 0.875 | 0.260 | 0.008 | 0.109 | 4.506 | 0.222 |
| EXP 2 | 21.1 | 87.700[a] | 64.300[a] | 56.200[a] | 107.900a | 0.518[b] | 10.670 | 398.500 | 76.470[a] |
| | Cycled[4] | 84.600[b] | 61.800[b] | 52.300[b] | 96.100[b] | 0.549[a] | 10.630 | 407.600 | 74.470[b] |
| | Pooled SE | 0.461 | 0.102 | 0.306 | 0.388 | 0.003 | 0.219 | 5.186 | 0.117 |

[a,b]Means within column with different superscripts were significantly different (p<0.05). [1-3]BBF, BAC, SWUSA defined as bone breaking force, bone ash content and shell weight unit surface area, respectively. [4]Cycled between 26.7 and 35.0

Table 5: Actual non-phytate P (NPP) and Ca intake for hens in experiment 1 (EXP 1)

| Temperature (°C) | Predicted NPP intake (mg h$^{-1}$ day$^{-1}$) | Actual NPP intake (mg h$^{-1}$ day$^{-1}$) | Predicted Ca intake (g h$^{-1}$ day$^{-1}$) | Actual Ca intake (g h$^{-1}$ day$^{-1}$) |
|---|---|---|---|---|
| 21.1 | 450 | 415 | 3.8 | 3.5 |
|  | 600 | 559 | 4.9 | 4.6 |
|  |  |  | 6.0 | 5.6 |
| 26.6 | 450 | 411 | 3.8 | 3.4 |
|  | 600 | 550 | 4.9 | 4.5 |
|  |  |  | 6.0 | 5.5 |
| 32.2 | 450 | 385 | 3.8 | 3.2 |
|  | 600 | 516 | 4.9 | 4.2 |
|  |  |  | 6.0 | 5.2 |

Table 6: Actual non-phytate P (NPP) intake for hens in Experiment 2 (EXP 2)

| Temperature (°C) | Predicted NPP intake (mg h$^{-1}$ day$^{-1}$) | Actual NPP intake (mg h$^{-1}$ day$^{-1}$) |
|---|---|---|
| 21.1 | 250 | 245 |
|  | 350 | 353 |
|  | 450 | 448 |
|  | 550 | 564 |
| Cycling[1] | 250 | 282 |
|  | 350 | 401 |
|  | 450 | 516 |
|  | 550 | 637 |

[1]Cycled between 26.7 and 35.0°C

Table 7: The effect of Ca dietary level on shell weight per unit surface area (SWUSA), bone breaking force (BBF) and bone ash concentration (BAC) (EXP 1)

| Ca dietary level (g) | BBF (kg) | BAC (mg/cm$^3$) | SWUSA (mg/cm$^2$) |
|---|---|---|---|
| 3.8 | 10.990[b] | 417.300[b] | 71.300[b] |
| 4.9 | 11.500[a] | 453.100[a] | 75.000[a] |
| 6.0 | 11.510[a] | 448.100[a] | 75.000[a] |
| Pooled SE | 0.109 | 4.506 | 0.222 |

[a,b]Means within column with different superscripts were significant different (p<0.05)

Table 8: The effect of environmental temperature and limestone *in vitro* solubility on shell weight per unit surface area (SWUSA) (EXP 1)[1]

|  | Environmental temperature (°C) | Limestone solubility (%) | SWUSA (mg/cm$^2$) |
|---|---|---|---|
| Exp 1 | 21.1 | 34.1 | 76.200[a] |
|  |  | 48.4 | 74.000[b] |
|  | 26.6 | 34.1 | 73.600[bc] |
|  |  | 48.4 | 72.400[c] |
|  | 32.2 | 34.1 | 72.300[c] |
|  |  | 48.4 | 73.600[bc] |
|  | Pooled SE |  | 0.222 |

[a,b,c]Means within column with different superscripts were significantly different (p<0.05). [1]The interaction effect of environmental temperature *limestone solubility* shell weight per unit surface area (SWUSA) was significant (p<0.01)

Table 9: The effect of environmental temperature and *in vitro* limestone solubility on shell weight per unit surface area (SWUSA) (EXP 2)[1]

| Temperature (°C) | Limestone solubility (%) | SWUSA (mg/cm$^2$) |
|---|---|---|
| 21.1 | 34.1 | 77.300 |
|  | 48.4 | 75.900 |
| Cycled[2] | 34.1 | 74.900 |
|  | 48.4 | 74.500 |
| Pooled SE |  | 0.117 |

[1]The interaction effect of environmental temperature *limestone solubility* shell weight per unit surface area (SWUSA) was significant (p<0.01). [2]Cycled between 26.7 and 35.0°C

4.9 and 6.0 g Ca groups (Table 7). The results indicate across all ET that laying hens have a higher Ca requirement for egg shell quality and bone integrity than 3.25 g hen$^{-1}$ day$^{-1}$ as suggested by NRC[3].

The significance (p<0.01) of the interaction effect of LS and temperature on SWUSA in both EXP 1(Table 8) and EXP 2 (Table 9) indicates that the optimum LS for shell quality is dependent on ET. The large particle limestone with low LS (34.1%) significantly increased the SWUSA value for egg shell quality for hens housed at 21.1°C in both EXP 1 and EXP 2. The large particle limestone with low LS (34.1%) did not significantly improve SWUSA for hens housed at 26.6°C and 32.2°C in EXP 1 and cycling hot ET in EXP 2. There was no significant interaction effect of LS and temperature on bone parameters (p>0.05). A number of researchers have demonstrated that large particle limestone with a low solubility has beneficial effects on egg shell and bone quality[18-21,24,38,41-50]. In normal thermoneutral temperatures, laying hens fed large particle limestone or oyster shell with lower LS have been shown to store multiple limestone or oyster shell particles in the crevices of the gizzard. The stored limestone or oyster shell particles are solubilized during the dark hours when the egg is in the shell gland providing additional Ca for egg shell formation. The reason is not clear for the lack of beneficial effect of lower LS on egg shell quality from hens housed in higher ET in the present study. Previous research has found that hens exposed to thermal ET increase their respiration rate causing reduced blood $pCO_2$[51] along with a reduction in blood bicarbonates[52]. Cheng *et al*.[53] reported that layers going through chronic heat stress with continuous 31.1°C and 60% relative humidity (RH) for a 12 week period show increasing respiratory alkalosis with increased blood pH and significant reductions in shell quality, Haugh units, $pCO_2$ and $HCO_3^-$ compared to layers housed in continuous 31.1°C ET with a lower heat index and a lower RH(40%). The LS and particle size was not evaluated in hot ET studies by Cheng et al.[53] but the authors used a 60:40 grandular : small particle blend of same limestone source and particle size as utilized in present studies for EXP 1 and 2. Odom et al.[54] reported that hens placed in acute hot ET immediately went into respiratory alkalosis with elevated pH. The respiratory alkalosis triggered the production of blood organic acids (lactate and pyruvate) that lowered the pH and reduced ionizable $Ca^{2+}$ available for eggshell formation by forming complexed calcium. The effects of limestone solubility on blood ionizable $Ca^{2+}$ levels for laying hens housed in hot ET has not been reported. Gordon and Roland[55] reported that environmental temperature did not affect in vivo solubility of limestone fed to laying hens by measuring non-solubilized limestone in the excreta. The researchers reported that environmental temperature did not affect in vivo limestone solubility when hens consumed equal Ca intake by evaluating calcium in excreta. The researchers found that hot ET slowed feed passage rate of the layers. The researchers did not evaluate the effect of hot ET on in vivo solubility in layers fed diets with different limestone particle sizes. Layers fed diets with large particle limestone in hot ET in EXP 1 and 2 in present study may not have been able to provide additional ionizable $Ca^{2+}$ that is reduced during respiratory alkalosis. Andrade et al.[33] reported that hens going through heat stress (diurnal or constant) do not produce improved shell quality by increasing dietary Ca intake. Franco-Jimenez and Beck[56] and Franco-Jimenez et al.[57] reported that hens housed in hot ET have a reduced Ca uptake in the intestine compared to hens fed same Ca in thermoneutral ET. The researchers evaluated acid base balance, endocrine status and Ca homeostasis in different strains in hot ET as well as thermotolerance with pre-exposure to hot ET. The researchers showed strain effects for different parameters in hot ET with the commercial Brown Egg strains being more susceptible. Ebeid et al.[58] reported that hens housed in high ambient temperatures have a reduced calbindin-D28k in intestinal segments and eggshell gland compared to calbindin-D28k in same tissue of hens housed in thermoneutral temperatures. Feeding a higher concentration of Ca or large particle limestone may be beneficial in thermoneutral temperatures for improving shell quality as shown in present study but the probable reduction of blood ionizable $Ca^{2+}$ for hens in hot ET could be caused by limited calbindin-D28k. The homeostatic acid-base response of hens to decrease pH during respiratory alkalosis by increasing the production of blood lactate and pyruvate may have limited the impact of additional Ca or large particle limestone to improve egg shell quality during heat stress. Manangi et al.[59] reported breeders fed diets containing small particle/highly soluble limestone produced a different blood ionizable $Ca^{2+}$ pattern during a 24 h post oviposition egg production cycle compared to breeders fed large particle/low soluble limestone. Breeders fed diets with large particle limestone produced a declining slope of plasma ionizable $Ca^{2+}$ from oviposition to 11-18 h after oviposition and the highest plasma ionizable $Ca^{2+}$ at 18-20 h. Mananga et al.[59] reported that urine pH values were significantly lower for breeders fed small particle limestone for the 6-11hr period after oviposition compared to urine collected at same time from breeders fed large particle limestone. The rationale for breeders producing a more acidic urine when fed controlled amounts of a breeder diet containing a smaller more soluble limestone particle is unknown.

Available P or digestible P requirements for commercial layers are primarily based on performance traits such as egg production, feed efficiency, feed intake, shell quality, mortality and maintenance or development of the skeletal system. Since layers are producing more egg mass and shell mass each year and the duration of production period is longer, concern with skeletal health and development of osteoporosis becomes more important. Bone status of layers tend to become more important in later stages of egg production because of cumulative depletion of bone P caused by daily bone mobilization of Ca for shell formation. Discrepancies exists for the suggested requirements of NPP or available P (AP) because of main criteria used in establishing requirements, age of layer, previous diets fed, acute or chronic high environmental temperatures (cycling or constant), vaccinations and immunological challenges, calcium levels and AP value given for ingredients and commercial phytase. Previously, Rama Rao et al.[60] conducted research on 5632 commercial White Leghorn layers from 22-72 week fed with diets containing 3.86% Ca with NPP levels ranging from 0.15-0.325%. The NPP requirement for egg production, feed intake, feed efficiency, shell quality and bone mineralization was 137.3 mg NPP from 22-37 week (phase I), 278.3 mg NPP from 37-54 week (phase II) and 262.0 mg NPP from 55-72 week (phase III)[60]. The researchers[60] determined that feed intake, feed efficiency and percent bone ash were affected by different dietary NPP levels in phase II and III whereas 0.15% dietary NPP was adequate for these traits during phase I and adequate for egg production and egg weight for all phases. White-Leghorn laying hens housed in ambient temperatures (experiment 1, 13-29.6°C and experiment 2; 8-17°C) and constant hot temperatures (experiment 1, 33°C and experiment 2, 35°C) fed diets with 3.5% Ca and increasing levels of NPP (experiment 1, 0.20-0.50% and experiment 2, 0.15-0.45%) have been reported to show an inverse relationship of dietary NPP with a decreased shell thickness[61]. Since shell quality is an important component in layer feeding, it becomes very easy to understand why formulations tend to have lower AP or NPP in post peak formulations. Layers in experiment 1, 22-34 week of age, fed diets with increasing NPP did not show an NPP response above 0.20% for feed conversions, body weight gain, feed intake, egg production, egg weight, egg mass and shell thickness. Post-peak layers in experiment 2: (1) Had the lowest body weight gain when fed 0.15% NPP diets and weight gain was significantly less than that of hens fed 0.45%, (2) Had the lowest feed intake when fed 0.15 and 0.25% NPP diets and was significantly less than that of hens fed 0.35 and 0.45% NPP and (3) Had the poorest feed conversions and egg production when fed 0.15% NPP and was significantly poorer than that of hens fed 0.35 and 0.45% NPP. Layers fed increasing NPP levels were reported to have increasing serum inorganic P and increasing percent tibia ash in experiment 2. In experiment 2, layers housed in the constant hot temperatures also had lower tibia ash and lower tibia P compared to hens housed in cooler temperature. The dietary inclusion of commercial phytase enzymes in layer diets has increased P efficiency and decreased P excretion due to better utilization of phytate P. Boling et al.[62] reported that layers housed in thermoneutral ET (20-70 week of age) consuming corn-soybean meal diets needed (0.15% AP) 159 mg AP day$^{-1}$ (without added phytase) or (0.10% AP plus 300 units phytase) 108 mg AP day$^{-1}$ to support optimum egg production for the long term study. The authors[62] suggested that additional P may be needed for older layers. The authors[62] reported that young layers could show optimum performance for as long as 13 week with only basal (0.10% AP) AP levels whereas older layers fed basal levels will start showing reduction in performance within 3 week. Coon and Leske[19] conducted a 16 week phytase study with 1200 commercial Hy-Line 26 week old W-36 layers housed in cages in thermoneutral temperatures and assorted into 24 different Ca and NPP treatments. Phytase response was affected by dietary Ca%, NPP% and limestone solubility (particle size) for egg production, egg mass, feed conversion, feed intake and weight gain during the study[19]. The higher response of phytase in improving performance, egg shell quality and tibia bone strength was achieved when hens were fed diets containing low soluble/large particle limestone having 3.5% Ca and 0.128% NPP as compared to feeding either high soluble or low soluble limestone having 0.228 and 0.328% NPP[19]. Phytase added to each of the NPP increments, except 0.428%, improved performance but some hens fed higher levels of NPP with phytase did not show significant results. The layers fed diets with and without phytase containing 3.5% Ca primarily from lower soluble limestone (3.3-4.7 mm particle size) with each of the test P levels showed an overall improvement in performance compared to hens fed comparable Ca and NPP levels with highly soluble limestone (0.5-0.8 mm particle size). The layers fed phytase with highly soluble Ca responded to phytase but the overall response of phytase was less. The lower solubility of Ca from larger limestone particles may interfere less during hydrolysis of phytate P. It is due to lesser formation of Ca phytate complex in the anterior part of GI tract. Coon and Leske[19] reported a slightly higher NPP requirement (0.205% AP, 0.128% AP +300 units phytase, 183 mg AP day$^{-1}$) for performance, shell quality and tibia bone strength for layers (26-42 week of age) compared to Boling et al.[62] Coon and Leske[19] reported increasing tibia breaking strength for hens fed increasing dietary AP (0.128-0.428% diet NPP; with or without 300 units of phytase added). The increasing dietary NPP increased tibia breaking strength for layers fed either limestone source[19]. Coon and Leske[19] indicated that lower NPP intake may support egg production and shell quality as compared to higher NPP intake that is required for maximum bone strength. According to Sell et al.[14], older hens fed with low NPP levels for extended periods may become susceptible to osteoporosis and mortality. In the present research, BBF and BAC was enhanced for hens fed the low LS in EXP 1 (p<0.05) (Table 10). Similar results with non-significant (p>0.05) differences were obtained in EXP 2. BBF and BAC both significantly increased with increasing dietary NPP intake (p<0.05) (Table 10) in EXP 1 and BAC significantly increased with increasing dietary NPP intake in EXP 2. The BBF also increased non significantly with increasing NPP intake in EXP 2. A significant interaction effect (p<0.05) of LS and daily NPP level on SWUSA was found in EXP 1 (Table 11). When lower LS (34.1%) was used, the higher NPP dietary intake (600 mg h$^{-1}$ day$^{-1}$) showed a detrimental effect on SWUSA but the NPP intake had no effect on SWUSA when hens were fed the limestone with a higher (48.4%) LS (Table 11). The interaction effect of LS and NPP intake on SWUSA was not significant in EXP 2. The SWUSA was higher for hens fed the limestone with 34.1% LS compared to the hens fed limestone with 48.4% LS in EXP 2 (Table 10). There were no significant differences in SWUSA with differences in NPP intake in EXP 2 (Table 10). In EXP 2, across both environmental temperatures,

Table 10: The effect of limestone *in vitro* solubility and dietary non-phytate P (NPP) intake on bone breaking force (BBF), bone ash concentration (BAC) and shell weight per unit surface area (SWUSA) in EXP 1 and EXP 2

|  | BBF (kg) | | BAC (mg/cm³) | | SWUSA (mg/cm²) |
| --- | --- | --- | --- | --- | --- |
|  | EXP 1 | EXP 2 | EXP 1 | EXP 2 | EXP 2 |
| **Limestone solubility (%)** | | | | | |
| 34.1 | 11.6[a] | 13.8 | 453.3[a] | 412.1 | 75.95[a] |
| 48.4 | 11.0[b] | 13.5 | 423.2[b] | 394.1 | 74.98[b] |
| **NPP intake (mg h⁻¹ day⁻¹)** | | | | | |
| 250 | - | 10.3 | - | 380.8[b] | 75.69 |
| 350 | - | 10.3 | - | 404.9[a b] | 75.68 |
| 450 | 11.0[b] | 10.9 | 429.1[b] | 412.2[b] | 75.68 |
| 550 | - | 11.2 | - | 414.5[a] | 75.61 |
| 600 | 11.6[a] | - | 447.9[a] | - | - |
| Pooled SE | | | | | 0.117 |

[a,b]Means within column with different superscripts were significant different (p<0.05)

Table 11: The effect of *in vitro* limestone solubility*dietary non-phytate P intake (NPP) on shell weight per unit surface area (SWUSA) (EXP 1)[1]

| Limestone solubility (%) | Dietary NPP intake (mg h⁻¹ day⁻¹) | SWUSA (mg/cm²) |
| --- | --- | --- |
| 34.1 | 450 | 74.500[a] |
| 34.1 | 600 | 72.700[b] |
| 48.4 | 450 | 73.700[ab] |
| 48.4 | 600 | 73.800[ab] |
| Pooled SE | | 0.117 |

[a,b]Means within column with different superscripts were significantly different (p<0.05). [1]The interaction effect of limestone solubility and NPP intake on shell weight per unit surface area was significant (p<0.05)

Table 12: Effect of environmental temperature*dietary Ca intake on shell weight per unit surface area (SWUSA) (EXP 1)[1]

| Environmental temperature (°C) | Dietary Ca intake (g h⁻¹ day⁻¹) | SWUSA (mg/cm²) |
| --- | --- | --- |
| 21.1 | 3.8 | 71.200[cd] |
| 21.1 | 4.9 | 77.500[a] |
| 21.1 | 6.0 | 76.800[ab] |
| 26.6 | 3.8 | 70.400[d] |
| 26.6 | 4.9 | 74.200[b] |
| 26.6 | 6.0 | 74.500[b] |
| 32.2 | 3.8 | 72.500[c] |
| 32.2 | 4.9 | 72.900[bc] |
| 32.2 | 6.0 | 73.500[bc] |
| SEM | | 0.674 |

[a,b,c,d]Means within column with different superscripts were significantly different (p<0.05). [1]The interaction effect of limestone solubility and NPP intake on shell weight per unit surface area was significant (p<0.05)

the layers needed a minimum intake of 350 mg NPP h⁻¹ day⁻¹ (AI = 353 mg at 21.1°C and 401 mg at the cycled hot temperature) (Table 6) for maximum BBF and BAC (Table 10). The present results indicate that a higher NPP intake is required for maintaining bone status compared to NPP intake required for supporting shell quality. The increased bone strength response for layers fed increasing dietary NPP would be expected because of the substantial amount of P incorporated into bone. Increasing the NPP intake in EXP 2 (across both environmental temperatures) did not increase or decrease SWUSA and shell quality (Table 10).

The highest SWUSA in EXP 1 was produced for hens fed with 4.9 g Ca dietary intake (4.4 g AI) and housed at 21.1°C (Table 5) and 6.0 g Ca dietary intake (5.4 g AI) (Table 5) was needed for highest SWUSA for hens housed at 26.6 and 32.2°C ET (Table 12). However, no significant differences were found between 4.9 and 6.0 g Ca intake groups when compared within each of the temperature groups (Table 12). The effect of LS on SWUSA was different with ET (p<0.05) as previously discussed for both EXP 1 (Table 8) and EXP 2 (Table 9). In EXP 1, low LS showed a beneficial effect on SWUSA when the limestone was fed to hens housed at 21.1°C but not at 26.6 and 32.2°C (Table 8). A similar effect was also observed in Exp 2 (Table 9). The best egg shell quality (SWUSA) for layers housed at 26.6°C ET was obtained when layers were fed 6.0 g Ca h⁻¹ day⁻¹ with lower LS (34.1% solubility) (Table 13). The best egg shell quality (SWUSA) for hens housed at 32.2°C was obtained by feeding 6 g Ca h⁻¹ day⁻¹ using the limestone with the highest LS (48.4% solubility) (Table 13).The limestone with the lowest LS (34.1% solubility) provides advantages for egg shell quality for layers housed at 21.1 and 26.6°C whereas layers housed at 32.2°C produced poorer egg shell quality when fed the limestone with lowest LS (34.1% solubility). The layers housed at 32.2°C produced the best egg shell when fed the limestone with the highest LS (48.4% solubility). The inability to improve egg shell quality for layers housed in hot ET by feeding a lower soluble/large particle limestone is probably related to decreased Ca transport in the intestine and respiratory alkalosis lowering blood $CO_2$, $HCO_3$ and ionizable $Ca^{2+}$. The Ca AI requirement for hens in EXP 1 is 4.2 g h⁻¹ day⁻¹ for optimum shell quality which supports the idea that older hens require a higher Ca intake than 3.75 g h⁻¹ day⁻¹ [39,40].

It is generally accepted that shell quality decreases with increasing level of NPP[46,47]. In EXP 1, the detrimental effect of increasing NPP dietary level on SWUSA was observed only for

Table 13: The effect of environmental temperature *limestone *in vitro* solubility* dietary Ca intake on shell weight per unit surface area (SWUSA) (EXP 1)[1]

| Temperature (°C) | Limestone solubility (%) | Dietary Ca intake (g h$^{-1}$ day$^{-1}$) | SWUSA (mg/cm$^2$) |
|---|---|---|---|
| 21.1 | 34.1 | 3.8 | 72.100 |
| 21.1 | 34.1 | 4.9 | 78.400 |
| 21.1 | 34.1 | 6.0 | 78.400 |
| 21.1 | 48.4 | 3.8 | 70.300 |
| 21.1 | 48.4 | 4.9 | 76.600 |
| 21.1 | 48.4 | 6.0 | 75.000 |
| 26.6 | 34.1 | 3.8 | 70.600 |
| 26.6 | 34.1 | 4.9 | 74.100 |
| 26.6 | 34.1 | 6.0 | 76.300 |
| 26.6 | 48.4 | 3.8 | 70.300 |
| 26.6 | 48.4 | 4.9 | 74.400 |
| 26.6 | 48.4 | 6.0 | 72.800 |
| 32.2 | 34.1 | 3.8 | 72.900 |
| 32.2 | 34.1 | 4.9 | 73.200 |
| 32.2 | 34.1 | 6.0 | 70.200 |
| 32.2 | 48.4 | 3.8 | 70.000 |
| 32.2 | 48.4 | 4.9 | 72.500 |
| 32.2 | 48.4 | 6.0 | 76.400 |
| Pooled SE | | | 0.222 |

[1]The interaction effect of environmental temperature *limestone solubility* dietary Ca intake on shell weight per unit surface area (SWUSA) was significant ($p<0.01$)

Table 14: The effect of environmental temperature *Ca intake* non-phytate P (NPP) intake on SWUSA (EXP 1)

| Environmental temperature (°C) | Dietary Ca intake (g h$^{-1}$ day$^{-1}$) | Dietary NPP intake (mg h$^{-1}$ day$^{-1}$) | SWUSA (mg/cm$^2$) |
|---|---|---|---|
| 21.1 | 3.8 | 450 | 71.600$^{bc}$ |
| 21.1 | 3.8 | 600 | 70.900$^{bc}$ |
| 21.1 | 4.9 | 450 | 78.700$^{a}$ |
| 21.1 | 4.9 | 600 | 75.900$^{ab}$ |
| 21.1 | 6.0 | 450 | 78.000$^{a}$ |
| 21.1 | 6.0 | 600 | 75.400$^{ab}$ |
| 26.6 | 3.8 | 450 | 71.000$^{bc}$ |
| 26.6 | 3.8 | 600 | 69.800$^{c}$ |
| 26.6 | 4.9 | 450 | 75.100$^{ab}$ |
| 26.6 | 4.9 | 600 | 73.200$^{b}$ |
| 26.6 | 6.0 | 450 | 73.700$^{b}$ |
| 26.6 | 6.0 | 600 | 75.300$^{ab}$ |
| 32.2 | 3.8 | 450 | 71.800$^{bc}$ |
| 32.2 | 3.8 | 600 | 73.100$^{bc}$ |
| 32.2 | 4.9 | 450 | 72.800$^{bc}$ |
| 32.2 | 4.9 | 600 | 73.000$^{bc}$ |
| 32.2 | 6.0 | 450 | 73.300$^{b}$ |
| 32.2 | 6.0 | 600 | 73.400$^{b}$ |
| SEM | | | 0.953 |

$^{a,b,c}$Means within column with different superscripts were significantly different ($p<0.05$). [1]The interaction effect of environmental temperature * Ca intake * NPP intake * shell weight per unit surface area (SWUSA) was significant ($p<0.05$)

hens housed at 21.1°C, not for those housed at 26.6 and 32.2°C. The NPP effect on SWUSA was also dependent on Ca intake ($p<0.05$) as observed in EXP 1 (Table 14). In EXP 2, small reductions in SWUSA were observed when NPP level was over 350 mg at 21.1°C, however there was no significant effect of dietary level of NPP on SWUSA for hens housed at either

Table 15: The effect of temperature and dietary non-phytate P (NPP) intake on shell weight per unit surface area (SWUSA) (EXP 2)

| Temperature (°C) | NPP intake (mg h$^{-1}$ day$^{-1}$) | SWUSA (mg/cm$^2$) |
|---|---|---|
| 21.1 | 250 | 76.62 |
| | 350 | 76.91 |
| | 450 | 76.55 |
| | 550 | 76.56 |
| Cycling[1] | 250 | 74.81 |
| | 350 | 74.44 |
| | 450 | 74.83 |
| | 550 | 74.63 |

[1]Cycled between 26.7 and 35.0°C

temperature(Table 15). In EXP 2, the hens received the same level of Ca with each NPP level for 21.1°C and hot cycling temperatures, whereas in EXP 1, the hens were fed three levels of Ca with two NPP levels at each temperature.

The NPP and Ca dietary levels had no significant effects on egg production in EXP 1 and 2. Furthermore, the different NPP levels had no significant affect on mortality in the hot ET. In EXP 1, mortalities (0.83, 0.83 and 4.2%) were recorded at 21.1, 26.6 and 32.2°C, respectively and in EXP 2, mortalities (1.3 and 2.1%) were recorded at 21.1°C and the cycling hot temperature. It was previously reported that insufficient dietary level of NPP depresses egg production and increases mortality[10-12]. The same effects were not observed in the present study. The duration of the experiments in the present study may not have been long enough to reproduce those effects if the body reserve of P was not depleted.

The body weights and weight gain were not significantly affected by Ca, NPP dietary levels and ET, although smaller weight gain was found for hens housed at 32.2°C compared with those housed at 21.1 and 26.6°C in EXP 1 (data not shown).

## CONCLUSION

Egg production decreased for commercial layers when ET was increased (32°C) compared to layers housed at 21.1 or 26.6°C. Egg production also decreased for layers housed in a more natural cycling ET (26.7 and 35.0°C) compared to a constant ET (21.1°C). The hot environmental temperatures reduced feed intake of the layers causing a reduction in egg number and egg weight. Egg shell quality decreased for the commercial White-Leghorn laying hen in both environmental nutrition studies as the temperature increased. Feeding older layers a diet containing large particle limestone with lower LS (34.1% *in vitro* solubility) improved egg shell quality for hens housed in thermoneutral ET (21.1°C). Feeding older layers a diet containing large particle limestone with lower LS (34.1% *in vitro* solubility) did not improve egg shell quality for hens housed at higher ET (constant or cycling).

Daily NPP intake of 245 and 353 mg h$^{-1}$ day$^{-1}$ supported optimum egg production and bone status at 21.1°C, respectively. Layers housed in hot ET (>30°C constant or 24 h mean cycling ET) may require 50 mg higher NPP intake h$^{-1}$ day$^{-1}$ to replace daily P loss due to mineral loss from endogenous depletion in hot ET compared with hens housed in cooler ET (21.1°C). Present research indicates that higher NPP may be required for bone status compared to egg production, especially in older hens. A minimum daily Ca intake (4.2 g) for both shell quality and bone status was needed for older hens ($\geq$48 week).

Research is needed on the physiological changes for laying hens housed in hot ET, especially those closely related with Ca and P utilization (for instance the pattern of HCl secretion of stomach, retention time of feed in the digestive tract, Ca and P mobilization from bone, P loss in urine) following mobilization from bone and impact on pH and buffering capacity. Also, additional research is needed to better understand *in vivo* limestone solubility when feeding different limestone particle sizes for layers housed in hot ET.

## ACKNOWLEDGMENT

The authors are indebted to ILC Resources, Des Moines, Iowa for partial financial support of the research and for providing limestone for the study. The authors are also indebted to discussions with Dr. Richard Miles (Emeritus University of Florida) regarding laying hen daily utilization of Ca and P in hot ET.

## REFERENCES

1. Keshavarz, K. and S. Nakajima, 1993. Re-evaluation of calcium and phosphorus requirements of laying hens for optimum performance and eggshell quality. Poult. Sci., 72: 144-153.
2. National Research Council, 1984. Nutrient Requirements of Poultry. 8th Rev. Edn., Washington, DC: The National Academies Press, United States, Pages: 80.
3. NRC., 1994. Nutritional Requirements of Poultry. 9th Edn., National Academy Science, Washington DC., USA.
4. Scheideler, S.E., 1994. Review of the basics of Ca and phosphorus nutrition in the laying hen. Proceedings of the Midwest Poultry Convention Egg Production Workshop, (MPCEP'94), Minnesota, USA.
5. Rostagno, H.S., L.F.T. Albino, M.I. Hannas, J.L. Donzele and N.K. Sakomura et al., 2017. Brazilian Tables for Poultry and Swine. Feedstuff Composition and Nutritional Requirements. 4th Edn., Universidade Federal de Viçosa, Viçosa, Brazil.
6. FEDNA., 2018. Nutritional Needs of Poultry. In: FEDNA Standards, 2nd Edn., Santoma, G. and G.G. Mateos (Eds.)., Madrid, Spain.
7. Rodriguez, M., W.J. Owings and J.L. Sell, 1984. Influence of phase feeding available phosphorus on egg production characteristics, carcass phosphorus content and serum inorganic phosphorus levels of three commercial layer strains. Poult. Sci., 63: 1553-1562.
8. Scheideler, S.E. and J.L. Sell, 1986. Effects of calcium and phase-feeding phosphorus on production traits and phosphorus retention in two strains of laying hens. Poult. Sci., 65: 2110-2119.
9. Said, N.W. and T.W. Sullivan, 1985. A comparison of continuous and phased levels of dietary phosphorus for commercial laying hens. Poult. Sci., 64: 1763-1771.
10. Singsen, E.P., A.H. Spandorf, L.D. Matterson, J.A. Serafin, J.J. Tlustohowicz, 1962. Phosphorus in the nutrition of the adult hen: 1. Minimum phosphorus requirements. Poult. Sci., 41: 1401-1414.
11. Singsen, E.P., L.D. Matterson, J.J. Tlustohowicz and W.J. Pudelkiewicz, 1969. Phosphorus in the nutrition of the adult hen: 2. The relative availability of phosphorus from several sources for caged layers. Poult. Sci., 48: 387-393.
12. Harms, R.H., B.L. Damron and R.D. Miles, 1977. Phosphorus and laying hen nutrition. Part II: Inadequate phosphorus levels result in increased mortality and decreased egg production. Feedstuffs, 49: 33-34.
13. Mikaelian, K.S. and J.L. Sell, 1981. Performance of laying hens fed various phosphorus levels continuously or phase fed decremental phosphorus levels. Poult. Sci., 60: 1916-1924.
14. Sell, J.L., S.E. Scheideler and B.E. Rahn, 1987. Influence of different phosphorus phase-feeding programs and dietary calcium level on performance and body phosphorus of laying hens. Poult. Sci., 66: 1524-1530.
15. Garlich, J.D., 1979. The phosphorus requirements of laying hens. Proceedings of Georgia Nutrition Conference, (GNC'79), Atlanta, GA., 104-114.
16. Charles, O.W., S. Duke and B. Reddy, 1978. Effect of phosphorus source and level on laying hen performance under varying temperature conditions. Georgia Nutr. Conf., 77: 74-87.
17. Belay, T., C.J. Wiernusz and R.G. Teeter, 1992. Mineral balance and urinary and fecal mineral excretion profile of broilers housed in thermoneutral and heat-distressed environments. Poult. Sci., 71: 1043-1047.
18. Cheng, T.K., R.L. Jevne and C.N. Coon, 1991. Calcium nutrition studies with layers. Proceedings of the 52nd Minnesota Nutrition Conference, September 16-18, 1991, Bloomington, MN., pp: 272-285.
19. Coon, C. and K. Leske, 1999. The effect of soluble calcium levels, nonphytate phosphorus levels and Natuphos phytase on laying hen performance. Proceedings of BASF Technical Symposium, Use of Natuphos Phytase in Layer Nutrition and Management, Atlanta, GA., (NPLNM'99), BASF Corporation, Mt. Olive, New Jersey, pp: 62-77.

*Int. J. Poult. Sci., 19 (5): 219-231, 2020*20. Coon, C., R. Ekmay and M. Manangi, 2011. Faecal, urinary and circulatory phosphorus flow in poultry. University of Minnesota. Minnesota Extension Service. Retrieved from the University of Minnesota Digital Conservancy, http://hdl handle.net/11299/204276.
21. Ekmay, R.D. and C.N. Coon, 2011. An examination of the P requirements of broiler breeders for performance, progeny quality and P balance 2. Ca particle size. Int. J. Poult. Sci., 10: 760-765.
22. Cheng, T.K. and C.N. Coon, 1990. Comparison of various *in vitro* methods for the determination of limestone solubility. Poult. Sci., 69: 2204-2208.
23. Zhang, B.F. and C.N. Coon, 1997. Improved *in vitro* methods for determining limestone and oyster shell solubility. J. Applied Poult. Res., 6: 94-99.
24. Cheng, T.K. and C.N. Coon, 1990. Effect of calcium source, particle size, limestone solubility *in vitro* and calcium intake level on layer bone status and performance. Poult. Sci., 69: 2214-2219.
25. Paganelli, C.V., A. Olszowka and A. Ar, 1974. The avian egg: Surface area, volume and density. Condor, 76: 319-325.
26. AOAC., 1990. Official Methods of Analysis: Association of Analytical Chemistry. 5th Edn., AOAC Inc., Washington, DC, USA.
27. Zhang, B. and C.N. Coon, 1997. The relationship of various tibia bone measurements in hens. Poult. Sci., 76: 1698-1701.
28. SAS., 2012. SAS User's Guide Statistics, Version 9.3. SAS Institute Inc., Cary, NC., USA.
29. De Andrade, A.N., J.C. Rogler and W.R. Featherston, 1976. Influence of constant elevated temperature and diet on egg production and shell quality. Poult. Sci., 55: 685-693.
30. Marsden, A. and T.R. Morris, 1987. Quantitative review of the effects of environmental temperature on food intake, egg output and energy balance in laying pullets. Br. Poult. Sci., 28: 693-704.
31. Oguntunji, A.O. and O.M. Alabi, 2010. Influence of high environmental temperature on egg production and shell quality: A review. World's Poult. Sci. J., 66: 739-750.
32. Payne, C.G., 1966. Practical aspects of environmental temperature for laying hens. World's Poult. Sci. J., 22: 126-139.
33. De Andrade, A.N., J.C. Rogler, W.R. Featherston and C.W. Alliston, 1977. Interrelationships between diet and elevated temperatures (Cyclic and Constant) on egg production and shell quality. Poult. Sci., 56: 1178-1188.
34. Vohra, P., W.O. Wilson and T.D. Siopes, 1979. Egg production, feed consumption and maintenance energy requirements of leghorn hens as influenced by dietary energy at temperatures of 15.6 and 26.7 C. Poult. Sci., 58: 674-680.
35. Tanor, M.A., S. Leeson and J.D. Summers, 1984. Effect of heat stress and diet composition on performance of white leghorn hens. Poult. Sci., 63: 304-310.
36. Alfredo, P. and C. Craig, 1991. Effect of temperature and dietary energy on layer performance. Poult. Sci., 70: 126-138.
37. Peguri, A. and C. Coon, 1993. Effect of feather coverage and temperature on layer performance. Poul. Sci., 72: 1318-1329.
38. Zollitsch, W., Z. Cao, A. Peguri, B. Zhang, T. Cheng and C. Coon, 1996. Nutrient requirements of laying hens. Proceedings of the International Symposium on Nutritional Requirements of Poultry and Swine, (ISNRPS'96), Departamento de Zootecnia, Vicosa-MG-Brasil, pp: 109-159.
39. Miller, P.C. and M.L. Sunde, 1975. The effects of precise constant and cyclic environments on shell quality and other lay performance factors with Leghorn pullets. Poult. Sci., 54: 36-46.
40. Wolfenson, D., Y.F. Feri, N. Snapir and A. Berman, 1979. Effect of diurnal or nocturnal heat stress on egg formation. Br. Poult. Sci., 20: 167-174.
41. Cufadar, Y., O. Olgun and A.O. Yildiz, 2011. The effect of dietary calcium concentration and particle size on performance, eggshell quality, bone mechanical properties and tibia mineral contents in moulted laying hens. Br. Poult. Sci., 52: 761-768.
42. Skrivan, M., M. Marounek, I. Bubancova and M. Podsednicek, 2010. Influence of limestone particle size on performance and egg quality in laying hens aged 24-36 weeks and 56-68 weeks. Anim. Feed Sci. Technol., 158: 110-114.
43. Pelicia, K., E. Garcia, C. Mori, A.B.G. Faitarone and A.P. Silva *et al.*, 2009. Calcium levels and limestone particle size in the diet of commercial layers at the end of the first production cycle. Revista Brasileira de Ciencia Avicola, 11: 87-94.
44. Zhang, B. and C.N. Coon, 1992. Nutrition Institute on Minerals. Chapter 7. Practical Applications. National Feed Ingredient Association, Chicago, IL.
45. Zhang, B., J.V. Caldas and C.N. Coon, 2017. Effect of dietary calcium intake and limestone solubility on egg shell quality and bone parameters for aged laying hens. Int. J. Poult. Sci., 16: 132-138.
46. Zhang, B., J. Weil, A.B. Guerra, P. Maharjan, K. Hilton, N. Suesuttajit and C.N. Coon, 2019. The effect of feeding various limestone particle sizes, limestone solubility and calcium intake on bone status and shell quality of a commercial white layer strain from 18-65 weeks of age. Int. J. Poult. Sci., 18: 372-378.
47. Cheng, T.K. and C.N. Coon, 1990. Sensitivity of various bone parameters of laying hens to different daily calcium intakes. Poult. Sci., 69: 2209-2213.
48. Scott, M.L., S.J. Hull and P.A. Mullenhoff, 1971. The calcium requirements of laying hens and effects of dietary oyster shell upon egg shell quality. Poult. Sci., 50: 1055-1063.
49. Brister, Jr. R.D., S.S. Linton and C.R. Creger, 1981. Effects of dietary calcium sources and particle size on laying hen performance. Poult. Sci., 60: 2648-2654.
230

50. Keshavarz, K. and C.C. McCormick, 1991. Effect of sodium aluminosilicate, oyster shell and their combinations on acid-base balance and eggshell quality. Poult. Sci., 70: 313-325.
51. El Hadi, H. and A.H. Sykes, 1982. Thermal panting and respiratory alkalosis in the laying hen. Br. Poult. Sci., 23: 49-57.
52. Barrett, N.W., K. Rowland, C.J. Schmidt, S.J. Lamont, M.F. Rothschild, C.M. Ashwell and M.E. Persia 2019. Effects of acute and chronic heat stress on the performance, egg quality, body temperature and blood gas parameters of laying hens. Poult. Sci., 98: 6684-6692.
53. Cheng, T.K., C.N. Coon and M.L. Hamre, 1990. Effect of environmental stress on the ascorbic acid requirement of laying hens. Poult. Sci., 69: 774-780.
54. Odom, T.W., P.C. Harrison and W.G. Bottje, 1986. Effects of thermal-induced respiratory alkalosis on blood ionized calcium levels in the domestic hen. Poult. Sci., 65: 570-573.
55. Gordon, R.W. and D.A. Roland, Sr., 1997. The influence of environmental temperature on *in vivo* limestone solubilization, feed passage rate and gastrointestinal pH in laying hens. Poult. Sci., 76: 683-688.
56. Franco-Jimenez, D.J. and M.M. Beck, 2007. Physiological changes to transient exposure to heat stress observed in laying hens. Poult. Sci., 86: 538-544.
57. Franco-Jimenez, D.J., S.E. Scheideler, R.J. Kittok, T.M. Brown-Brandl, L.R. Robeson, H. Taira and M.M. Beck, 2007. Differential effects of heat stress in three strains of laying hens. J. Applied Poult. Res., 16: 628-634.
58. Ebeid, T.A., T. Suzuki and T. Sugiyama, 2012. High ambient temperature influences eggshell quality and calbindin-D28k localization of eggshell gland and all intestinal segments of laying hens. Poult. Sci., 91: 2282-2287.
59. Manangi, M.K., P. Maharjan and C.N. Coon, 2018. Calcium particle size effects on plasma, excreta and urinary Ca and P changes in broiler breeder hens. Poult. Sci., 97: 2798-2806.
60. Rama Rao, S.V., M.V.L.N. Raju, S.S. Paul and B. Prakash, 2019. Effect of supplementing graded concentrations of non-phytate phosphorus on performance, egg quality and bone mineral variables in white leghorn layers. Br. Poult. Sci., 60: 56-63.
61. Usayran, N., M.T. Farran, H.H. Awadallah, I.R. Al-Hawi, R.J. Asmar and V.M. Ashkarian, 2001. Effects of added dietary fat and phosphorus on the performance and egg quality of laying hens subjected to a constant high environmental temperature. Poult. Sci., 80: 1695-1701.
62. Boling, S.D., M.W. Douglas, M.L. Johnson, X. Wang, C.M. Parsons, K.W. Koelkebeck and R.A. Zimmermant, 2000. The effects of dietary available phosphorus levels and phytase on performance of young and older laying hens. Poult. Sci., 79: 224-230.